\documentclass[oldversion]{aa} 
\usepackage{amsmath}

\usepackage{natbib}
\usepackage{amssymb}
\usepackage{graphicx}
\usepackage{txfonts}
\begin{document}
   \title{HST FUV imaging of DG~Tau}
   \subtitle{Fluorescent molecular hydrogen emission from the wide opening-angle outflow}

   \author{P.C. Schneider\inst{1}
          \and
          J. Eisl\"offel\inst{2}
          \and
          M. G\"udel\inst{3}
          \and
          H.M. G\"unther\inst{4}
          \and
          G. Herczeg\inst{5}
          \and
          J. Robrade\inst{1}
          \and
          J.H.M.M. Schmitt\inst{1}
          }

   \institute{Hamburger Sternwarte,
              Gojenbergsweg 112, 21029 Hamburg\\
              \email{cschneider@hs.uni-hamburg.de}
              \and
              Th\"uringer Landessternwarte Tautenburg, Sternwarte 5, 07778 Tautenburg, Germany
              \and
              Universit\"at Wien, Dr.-Karl-Lueger-Ring 1, 1010 Wien, Austria
              \and
              Harvard-Smithsonian Center for Astrophysics, 60 Garden Street, Cambridge, MA 02138, USA
              \and
              The Kavli Institute for Astronomy and Astrophysics, Peking University, Yi He Yuan Lu 5, Hai Dian Qu, Beijing 100871, China
             }

   \date{Received .. / accepted ..}

  \abstract
   {One of the best-studied jets from all young stellar objects is the jet of DG~Tau, which we imaged in the far-ultraviolet with the Hubble Space Telescope for the first time. These high spatial resolution images were obtained with long-pass filters and allow us to construct images tracing mainly $H_2$ and C~{\sc iv} emission. We find that the $H_2$ emission appears as a limb-brightened cone with additional emission close to the jet axis. The length of the rims is about 0\farcs3 or 42\,AU before their brightness strongly drops, and the opening angle is \hbox{about 90$^\circ$}. Comparing our FUV data with near-infrared data we find that the fluorescent $H_2$ emission probably traces the outer, cooler part of the disk wind  while an origin of the $H_2$ emission in the surface layers (atmosphere) of the (flared) disk is unlikely. Furthermore, the spatial shape of the $H_2$ emission shows little variation over six years which suggests that the outer part of the disk wind is rather stable and probably not associated with the formation of individual knots. The C~{\sc iv} image shows that the emission is concentrated towards the jet axis. We find no indications for additional C~{\sc iv} emission at larger distances, which strengthens the association with the X-ray emission observed to originate within the DG~Tau jet.
   }

   \maketitle
%

\section{Introduction}
Stars form by  gravitational collapse in molecular clouds. Soon after the initial radial infall ceases, they continue their accretion from a just formed circumstellar disk. This accretion process is accompanied by strong outflow activity. Single classical T~Tauri stars (CTTSs) are
of particular interest for the study of jets from young stellar objects (YSOs), because their relatively cleared environment allows the jets to be followed almost all the way down to their sources.

The single CTTS DG~Tau is the source of one of the first known outflows from a CTTS \citep[HH158, ][]{Mundt_1983}, and is still among the brightest optical/near-IR sources so that its properties have been studied by numerous observations \citep[e.g.,][]{Mundt_1987, Eisloeffel_1998, Coffey_2008, Agra_Amboage_2011, Rodriguez_2012}. 
The velocity structure of the DG~Tau jet is onion-like with high-velocity parts enclosed by slower material \citep{Bacciotti_2000}. There are also indications of jet rotation \citep{Bacciotti_2002, Coffey_2007} in the same sense as the disk \citep{Testi_2002}. The jet rotation and onion-like velocity structure are best explained by a disk origin of the outflow \citep{Anderson_2003, Ferreira_2006}. Whether there are additional contributions from a stellar wind \citep[e.g.,][]{Melani_2006} or an X-wind launched in the region, where the magnetic fields of the star  and the disk interact \citep{Shang_2002}, is currently unclear.

Most studies of the DG~Tau jet (and YSO jets in general) utilize bright optical and near-IR forbidden emission lines, e.g., [O~{\sc i}]\,$\lambda6300$, [S~{\sc ii}]\,$\lambda\lambda6716,6731$, or [Fe~{\sc ii}] at 1.64$\,\mu$m, which trace plasma temperatures around $10^4$\,K. These lines originate in the cooling zones of shocks within the jet; for DG~Tau shock-velocities of 50 to 100\,km\,s$^{-1}$ have been estimated \citep{Lavalley_Fouquet_2000}. 
In addition to the $10^4\,$K plasma, YSO jets,  and the jet of DG~Tau in particular, contain plasma of higher {\it and} lower temperature. Here, we use far-ultraviolet (FUV) observations to trace the low temperature part with molecular hydrogen ($H_2$) emission of material with only about $2\times10^3$\,K and C~{\sc iv} emission ($T\sim10^5$) to study the higher temperature regime. Since DG~Tau is essentially invisible in the FUV shortwards of about $1600$\,\AA{},  we can follow the $H_2$ and C~{\sc iv} emission extremely close to the star.

\begin{table*}[t]
\begin{minipage}[h]{0.99\textwidth}
\renewcommand{\footnoterule}{}
  
  \caption{Analyzed observations \label{tab:obs} }
  \begin{tabular}{c c c c r c r} \hline \hline
  Observing date & Instrument & Detector &  Filter/Grating & Exposure time & Wavelength range (\AA) & Obs. ID.\\     
  \hline

  1981 Oct 20 &  IUE & \multicolumn{2}{c}{Low dispersion, short wvl.}&  23.1\,ks & 1150 -- 1979 &  SWP15301\\
  1982 Jan 12& IUE & \multicolumn{2}{c}{Low dispersion, short wvl.} & 26.4\,ks & 1150 -- 1979 & SWP16033    \\
  1996 Feb 08 & HST & GHRS & G160 & 16.5\,ks & 1383 -- 1417 & 5875 \\
  1999 Jan 14 & HST & STIS & G750M & 16.5\,ks & 6300 -- 6870 & 7311 \\
  2000 Oct 22 & HST & STIS & E140M & 2.7\,ks & 1123 -- 1710 & 8157\\
  2011 Feb 17 & HST  & STIS FUV-MAMA & G140M & 14.0\,ks & 1514 -- 1567 & 12199\\
  2011 Jul 23 & HST   & ACS SBC & F140LP & 1.9\,ks & $1350$ -- $2000$& 12199\\
  2011 Jul 23 & HST  &  ACS SBC & F150LP & 1.9\,ks & $1450$ -- $2000$& 12199\\
  2011 Jul 23 & HST  &  ACS SBC & F165LP & 1.9\,ks & $1600$ -- $2000$& 12199\\
  \hline
  \end{tabular}
  \end{minipage}
\end{table*}

Far-ultraviolet molecular hydrogen emission results from electronic transitions which in general require FUV pumping by Ly$\alpha$ photons, while the ro-vibrational near-IR $H_2$ lines can be directly stimulated by shocks without FUV pumping \citep{Wolfire_1991}.
Fluorescent $H_2$ in the T~Tau system has been studied by \citet{Walter_2003} and \citet{Saucedo_2003}. In this system, the on-source $H_2$ emission is pumped by broad, self-absorbed Ly$\alpha$ emission which is related to the accretion process. At larger distances from T~Tau~N, all detected $H_2$ lines result from fluorescence routes which have pumping transitions close to the rest-velocity of Ly$\alpha$. Therefore, \citet{Walter_2003} concluded that the extended $H_2$ emission is not pumped by stellar Ly$\alpha$ emission but by emission from local shocks similar to the case in some Herbig-Haro (HH) objects. This interpretation differs from \citet{Saucedo_2003} who hypothesize stellar emission as the pumping source while, on the other hand, also requiring shock-heating to achieve the plasma temperature of $T>1000\,$K required for efficient Ly$\alpha$ fluorescence pumping.

Molecular hydrogen has been observed from the region around DG~Tau in the FUV \citep{Ardila_2002,Herczeg_2006, Schneider_FUV} and in the near-IR \citep{Takami_2004, Beck_2008}. The FUV and near-IR $H_2$ emissions are both concentrated in the forward jet direction. \citet{Takami_2004} calculated the momentum flux of the $H_2$ emitting material for excitation by stellar UV and X-ray photons and found that this momentum flux would be uncomfortably large for a class~II source like DG~Tau. Therefore, they favor shock heating for the near-IR $H_2$ lines.

Molecules in YSO outflows can originate from entrainment of ambient molecular, gas-phase formation in a dust-free atomic wind, or from a dusty disk wind (including $H_2$) with possible $H_2$ reformation. 
The spatial structure of the $H_2$ emission will constrain its origin as one expects the molecular part of a dusty disk wind located outside the atomic region while, for example, $H_2$ reformed behind shocks would be preferentially located in the path of the optical jet.
Several authors found that blue-shifted fluorescent $H_2$ emission is probably associated with outflows \citep[][]{Ardila_2002, Herczeg_2006, France_2012}. A prime example for this interpretation is RW~Aur due to the large velocity-shift of the $H_2$ emission.

The high-temperature part of jets ($T\gtrsim10^6$\,K) can be traced by X-ray emission. The jet of DG~Tau was the first CTTS jet discovered in X-rays \citep{Guedel_2005, Guedel_2007}. The favorable viewing geometry with a jet inclination close to 45$^\circ$ allowed \citet{Schneider_2008} to separate the hard coronal X-ray component from a softer X-ray component suspected to originate in the jet. The mean distance of the soft X-ray photons from DG~Tau  stays constant within six years \citep{Guedel_2011} and thus contrasts the optical forbidden emission lines (FELs), where proper-motion is clearly present \citep{Eisloeffel_1998,Dougados_2000}. 
The intermediate temperature region has been studied with Hubble Space Telescope (HST) Space Telescope Imaging Spectrograph (STIS) long-slit data of FUV C~{\sc iv} emission (peak formation temperature $10^5$\,K), showing that the $\sim10^5\,$K plasma is almost co-spatial with the stationary X-ray emission where the $H_2$ emission has a minimum \citep{Schneider_FUV}. This FUV spectrum is, however, limited to the region close to the jet axis owing to the slit-width of 0\farcs2, and the images presented here cover a much larger region around DG~Tau (Field of view (FOV): $32\times32$\,arcsec).

Our paper is structured as follows.  We  describe the observations and the data reduction including a description of the  $H_2$ and C~{\sc iv} image  construction in Sect.~\ref{sect:obs}. Results and discussion of $H_2$ (Sects.~\ref{sect:H2results} and \ref{sect:H2discussion}), C~{\sc iv} emission (Sect.~\ref{sect:c2}), and the outer knots (Sect.~\ref{sect:c3}) are divided into individual sections. We close with   a summary and conclusions in Sect.~\ref{sect:sum}.

\section{Observations and data reduction \label{sect:obs}}

\begin{figure}
\vspace*{0.1cm}
\centering
\fbox{\includegraphics[width=0.48\textwidth]{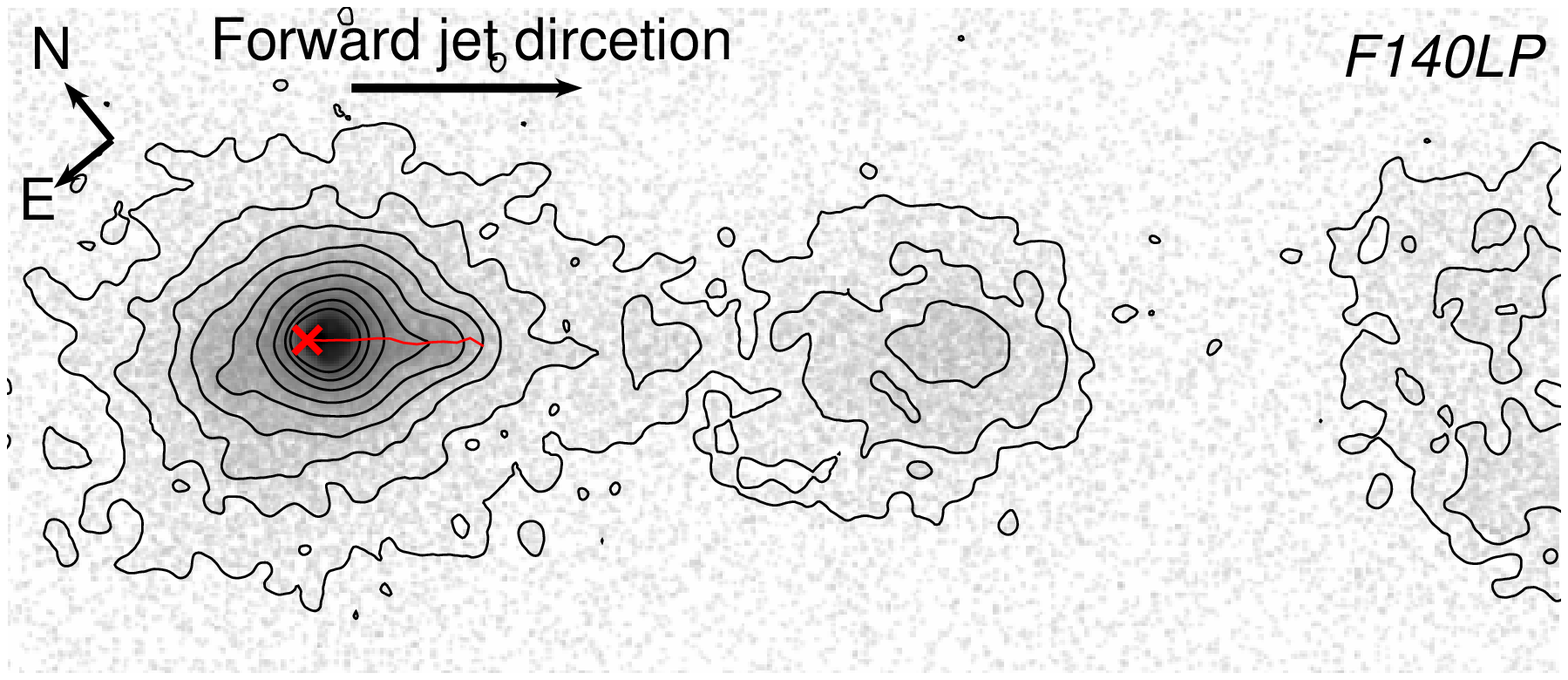}}
\fbox{\includegraphics[width=0.48\textwidth]{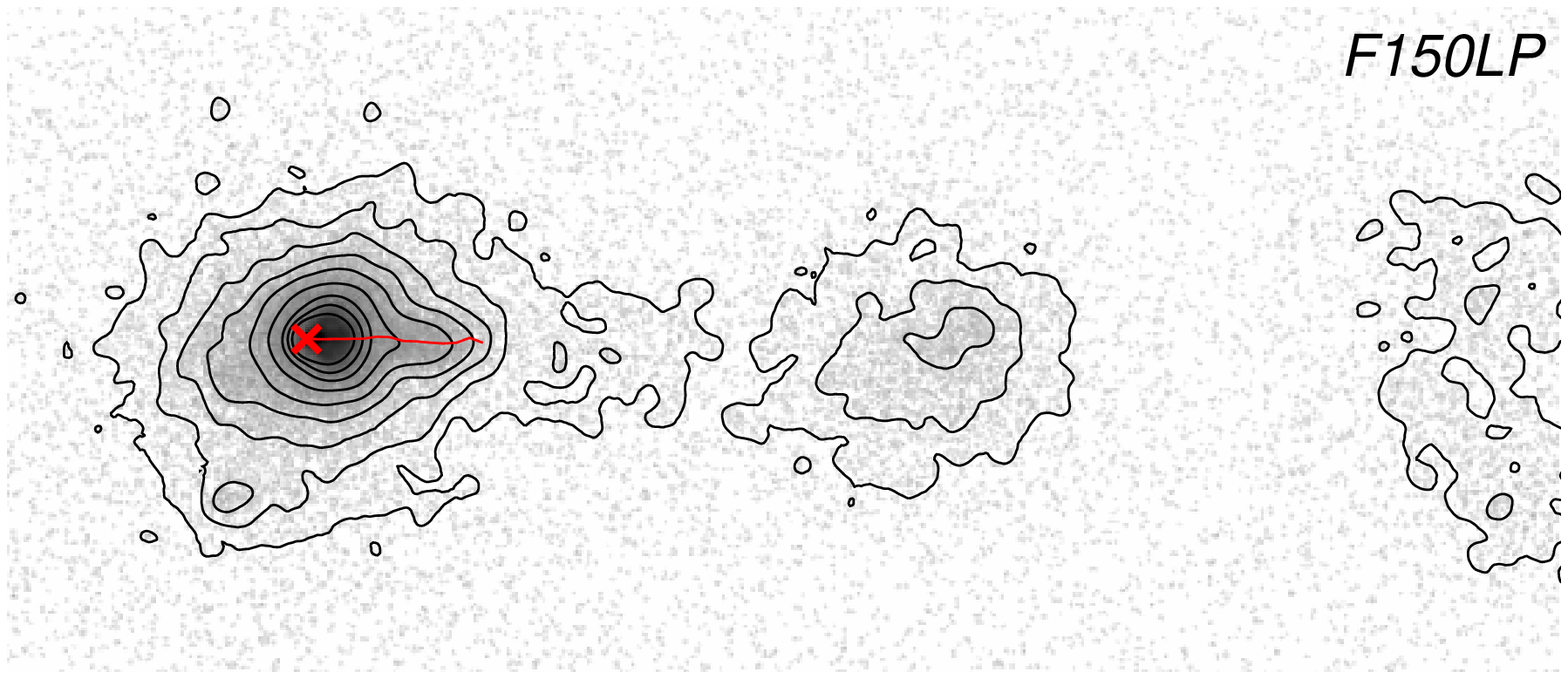}}
\fbox{\includegraphics[width=0.48\textwidth]{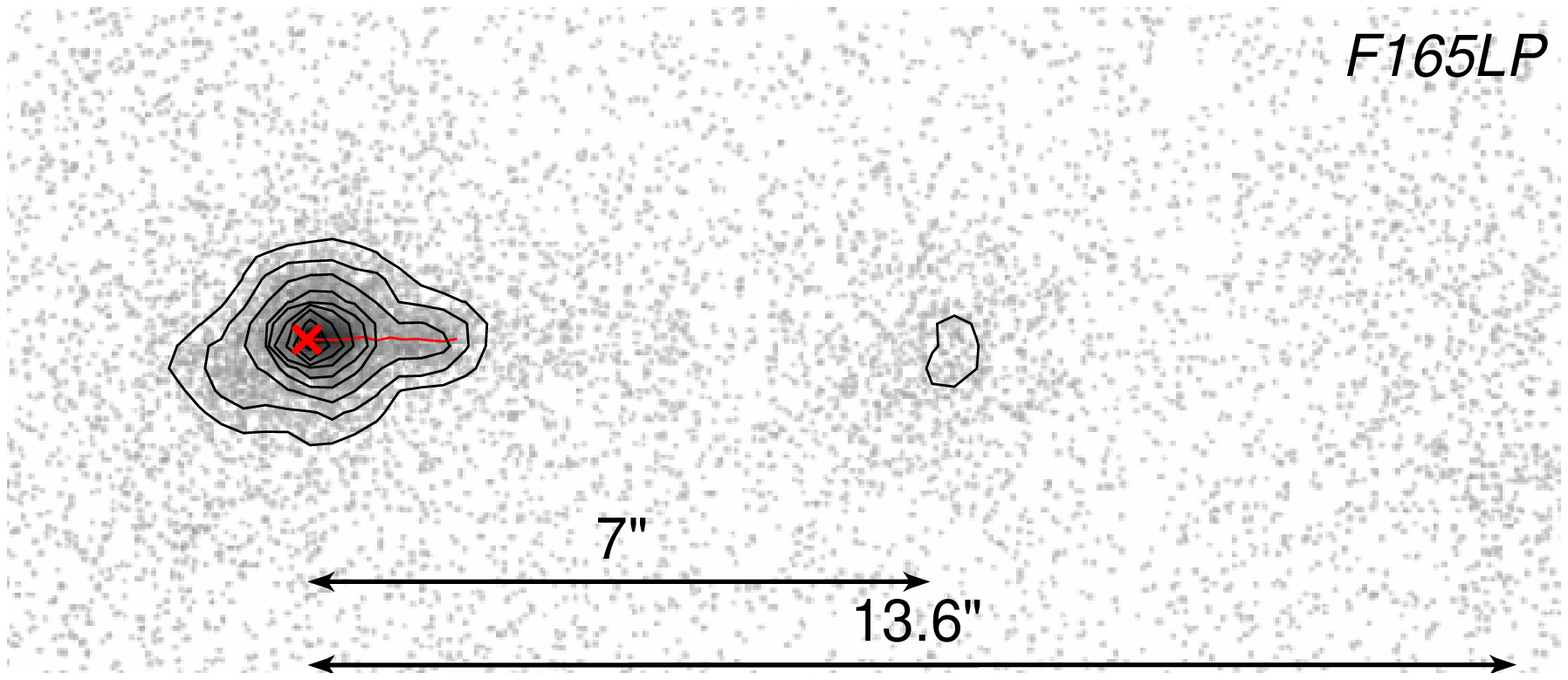}}
\caption{FUV images of the DG Tau jet. From top to bottom: F140LP, F150LP, F165LP. The ranges of the logarithmic scaling differ between panels. Contour levels are equal in all frames, they start at 2 counts/pix (smoothed by a Gaussian of 0\farcs25) and increase in steps of a factor of two. The cross indicates the position of DG~Tau (the size of the cross gives the uncertainty).The outermost emission knot is at the edge of the detector, and its bow/apex is outside the field of view. The red line indicates the jet center.\label{fig:rawIm} } 
\end{figure}

\begin{figure*}
\centering
\includegraphics[width=0.99\textwidth]{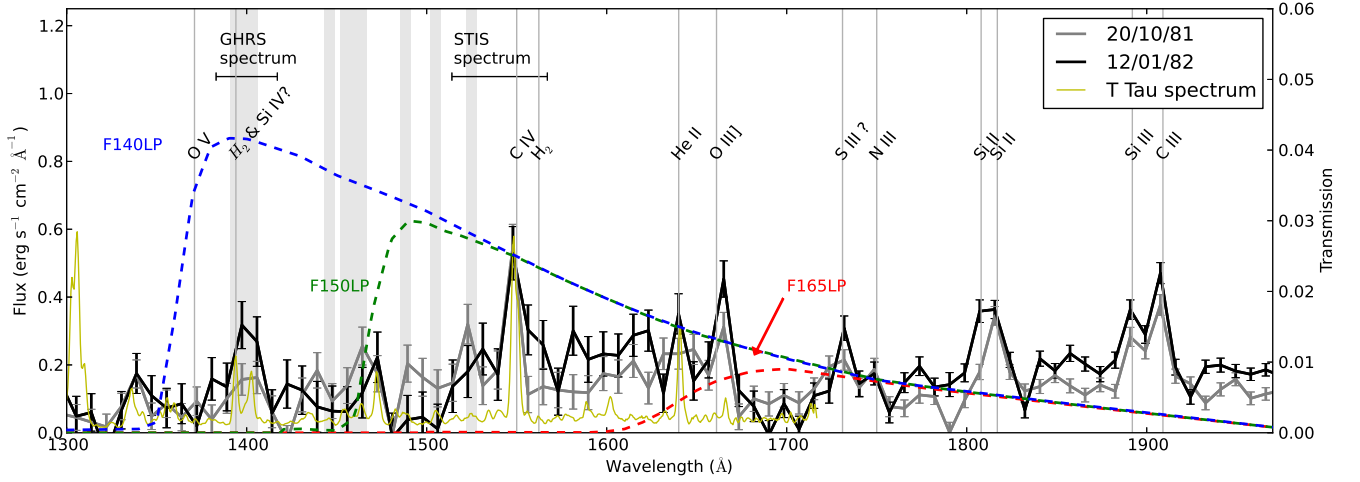}
\caption{Re-binned IUE FUV spectrum of DG Tau with filter transmission curves of the ACS images superimposed (right axis). Strong lines are labeled and the band-passes of the high-resolution spectra are indicated. The filter efficiencies are indicated by the dashed lines (right axis). For comparison, the higher resolution HST STIS spectrum of T~Tau is also shown (yellow). The gray-shaded regions indicate strong $H_2$ lines. \label{fig:IUE}}
\end{figure*}

\subsection{Observations}
In addition to our new HST FUV images we use a number of archival FUV observations of DG~Tau. Specifically, we use International Ultraviolet Explorer (IUE) spectra to constrain the broadband FUV spectrum (observation dates 1981 and 1982), high resolution FUV spectra obtained with the Goddard High Resolution Spectrograph (GHRS) onboard HST covering about 1\farcs74 around DG~Tau (observation date 1996), high resolution FUV Echelle spectra obtained with HST STIS covering the inner region around DG~Tau (0\farcs1$\times$0\farcs2, observation date 2000), and FUV long-slit spectra covering the region around the jet axis (observation date 2011). For the morphological comparison of our FUV images with the low temperature plasma seen in bright forbidden emission lines, we use the reconstructed image obtained from seven consecutive long-slit optical HST spectra obtained in 1999.

Our new DG Tau FUV images were taken with the Advanced Camera for Surveys (ACS) using the Solar Blind Channel (SBC). We obtained images in three long band-pass filters (F140LP, F150LP, and F165LP) within two consecutive HST orbits (see Table~\ref{tab:obs}). Two separate images were obtained in each filter and co-added for analysis. 
To minimize the relative offsets between the two images obtained with the same filter, they are registered using    \texttt{tweakshifts} from the PyRAF \texttt{multidrizzle}-package prior to co-adding them.  
The relative offsets for the exposures taken within the same orbit are  negligible ($\lesssim 0.1$\,pixel); only the two    {F150LP} images separated by an earth occultation have a relative offset of 0.4\,pixel ($12\,$mas).  

Figure \ref{fig:rawIm} shows the three FUV images. The position angle (PA) of the jet is about 225\,$^\circ$, which is not ideal for displaying purposes, so all images were rotated by 45$^\circ$, i.e., the forward jet now faces towards the right.
We clearly detect emission from the immediate region around DG~Tau and from two outer knots at about $7\arcsec$ and $14\arcsec$.  The background estimated from source-free circular regions with $r=2\farcs5$ is small compared to the emission. The signal to background is $>$ 4 (20) within the innermost 1\farcs2 (0\farcs5).  
Background subtraction was performed using the mean obtained from three $r=2\farcs5$ circles. Spatial fluctuations of the background are less than 20\,\%.

The longpass filters have overlapping throughputs longwards of their cut-off wavelengths, i.e., the  {F140LP} image includes all the emission seen in the    {F150LP}.  The throughputs\footnote{ \texttt{http://www.stsci.edu/hst/acs/analysis/throughputs}} are shown in Fig.~\ref{fig:IUE} overplotted onto the archival IUE spectra of DG~Tau.
Additionally, photons with wavelength around 3000\,\AA{} increase the count rate due to the red leak of the ACS SBC. We constructed a broadband spectrum using IUE data complemented by STIS G430L data scaled to smoothly match the IUE spectrum and find that up to 17\,\%  of the counts in the {F165LP} images might result from  photons with $\lambda>1900\,\AA$ (lower fractions pertain to the other filters). However, those photons are effectively removed during the following procedures.

\subsection{The FUV spectrum of DG Tau \label{sect:origin}}
Our aim is to extract images in $H_2$ and C~{\sc iv} from three long-pass filter images (F140LP, F150LP, and F165LP\footnote{The filter names indicate their short wavelength cutoff in nm.}) using difference images. 
Therefore, we have to analyze the spectral origin of the detected photons. Since their wavelengths are between about $1350$ and $2000$\,\AA{}, only
two archival IUE spectra cover the whole bandpass (we omit the short exposure from 1981 Jul 30, see Table~\ref{tab:obs} for observational details).  These spectra (shown in Fig.~\ref{fig:IUE}) were obtained in 1981 and 1982. They have a FOV of about $10\arcsec\times20\arcsec$ so that they contain emission from DG~Tau as well as from its jet. The two IUE spectra differ slightly in some wavelength regions, but most features are visible in both exposures.

In the wavelength range relevant for our HST FUV images, the spectra of CTTS contain atomic emission lines like C~{\sc iv}, molecular hydrogen emission lines, and a weak continuum \citep[sometimes with additional emission from ro-vibrational bands of CO, e.g.,][]{Schindhelm_2012}. Our HST G140M spectrum of DG~Tau and its jet around the C~{\sc iv} doublet \citep[$\lambda = 1514 - 1567$\,\AA,][]{Schneider_FUV} does not show any hint of continuum or CO emission. Therefore, the part of the  DG~Tau FUV spectrum relevant for our images is dominated by atomic  and $H_2$ emission lines  without any significant continuum contribution.

For the atomic emission, only a few lines can rival the strong C~{\sc iv} doublet at about $1550\,$\,\AA{} and  might contribute significantly to our images (essentially lines from Si~{\sc iv}, He {\sc ii},  [O {\sc iii}], and O~{\sc iv}]). 
An estimate of their contributions can be obtained from the two IUE spectra by integrating the flux within $\pm6$\,\AA{} around the nominal position of these lines (this region includes about $95\,\%$ of the line flux for a line shift below 300\,km\,s$^{-1}$). 
Since these regions also include strong $H_2$ lines, we use an archival HST GHRS spectrum \citep[see Table~\ref{tab:obs} and ][]{Ardila_2002} to further constrain the emission from Si~{\sc iv} and O~{\sc iv} lines \hbox{around 1400\,\AA{}.} This spectrum covers a sufficiently large region around DG~Tau to include jet emission (FOV: $1\farcs74\times1\farcs74$). 
It demonstrates that Si~{\sc iv} and O~{\sc iv} emission are at most marginally present since all emission lines consistently match with $H_2$ lines \citep[see also ][]{Ardila_2002}. We derive an upper limit of $\lesssim4.5\times10^{-15}$\,erg\,s$^{-1}$ from fitting  a set of Gaussians to the emission lines in the GHRS spectrum. No high spectral resolution data of the He {\sc ii} and [O {\sc iii}] lines covering DG~Tau {\it and} the jet exist so that we cannot improve the estimate based on the IUE spectrum for these lines. 
Their contributions are given in Table~\ref{tab:fractions}. 
The total luminosity excluding the wavelength regions of strong atomic emission lines is about $3.5\times10^{-13}$\,erg\,s$^{-1}$ (1300 -- 1650\,\AA) and is almost exclusively caused by $H_2$ emission.

\begin{table}[t]
\caption{Contribution of strong atomic lines to the observed counts in the difference images estimated from the IUE. Values in brackets are derived from the STIS E140H or GHRS spectra. The filters are abbreviated as F plus cutoff wavelength. }
\label{tab:fractions}
\begin{tabular}{l c c l r}
\hline\hline
Line & Wavelength  & Flux & \multicolumn{2}{c}{Count fraction in}\\
     &   (\AA)      & $10^{-15}$\,erg\,s$^{-1}$ &  \multicolumn{2}{c}{difference image } \\
\hline
Si {\sc iv}\tablefootmark{a} & 1388 - 1409 & 38 ($\lesssim4.5$) & F140 - F150: & 30\,\% (4\,\%) \\
C {\sc iv} & 1542 - 1557 & 55 (20)  & F150 - F165: & 18\,\% (9\,\%)\\
 He {\sc ii} & 1634 - 1646 & 29 & F150 - F165:& 5\,\% (6\,\%)\\
{}[O {\sc iii}] & 1657 - 1669 & 36 & F150 - F165:& 3\,\% (3\,\%)\\
\hline
\end{tabular}
\tablefoot{
\tablefoottext{a}{Includes emission from O {\sc iv}].}}
\end{table}

The observed count-ratios of 1:0.67:0.16 (F140LP:F150LP:F165LP) are close to the expectation from the IUE spectra (1:0.71:0.21) using the counts within a circle of 0\farcs3 around the brightest part of the emission. The observed count-rates are slightly higher than expected from the IUE spectrum; the largest discrepancies (a factor of \hbox{about $1.3$}) pertain to the F140LP filter, which might be due to cross-calibration issues and/or minor source-variability.
These two facts suggest that today's overall spectrum is comparable to the IUE data, which means that we can use the estimates obtained from the IUE and GHRS spectra to identify the dominating contributions to the filter images.

\subsection{Spectral contributions to the filter images}
Using specific combinations of the filter images, we can construct images tracing mainly $H_2$ and C~{\sc iv} emission. 
Convolving the IUE spectra with the detector response, we find that strong atomic emission lines contribute $\lesssim4$\,\% to the  \hbox{F140LP - F150LP} difference image.
Furthermore, an $H_2$ model based on branching ratios and the strongest $H_2$ lines detected in the E140M spectrum already reproduces $\gtrsim80\,$\% of the flux below 1500\,\AA. Therefore, the {F140LP - F150LP} image is largely dominated by $H_2$ emission and we estimate that atomic emission lines contribute only marginally to the recorded signal.

Within the  {F150LP - F165LP} difference image, strong atomic lines contribute $\lesssim26$\,\% of the observed counts. Most of the atomic emission in this spectral region is C~{\sc iv} emission \hbox{(about $18$\,\%)} with some additional contribution from He {\sc ii} and [O {\sc iii}] ($\lesssim 8\,$\%  of the counts).
Thus the {F150LP - F165LP} difference image is, like the  {F140LP - F150LP} image, again dominated by $H_2$ emission as spectra around 1550\,\AA{} show no continuum emission.
To extract the C~{\sc iv} emission (representative of any non-$H_2$ emission), the $H_2$ emission, which contributes most of the counts, must be removed.  Convolving the IUE spectra with the ACS/SBC filter responses and summing the wavelength-regions without the strong atomic lines mentioned above,
we find that {F140LP - F150LP} image contains 70\,\% of the $H_2$ counts of the    {F150LP - F165LP} image\footnote{From our recent STIS long-slit spectrum we know that about 85\,\% of the flux within 1542 -- 1557\,\AA{} is actually C~{\sc iv} emission and we corrected the fraction accordingly. }. A similar value is obtained from the HST STIS E140H spectrum \citep{Herczeg_2006}, excluding the same wavelength regions although this spectrum covers only the innermost 0\farcs1 of the jet, i.e.,  only a fraction of the $H_2$ emission. These values differ from an $H_2$ model based on branching ratios using the $H_2$ lines measured in the E140H spectrum.
Within the F150LP - F140LP wavelength  range, the predicted total flux of this branching ratio model approximately matches  the observed total flux, but falls short of the measured flux longwards of about $1500$\,\AA, i.e., it misses lines. Therefore, we prefer the results obtained by convolving the observed flux with the detector responses.

Assuming that the spatial  distribution of the $H_2$ emission traced by the {F140LP - F150LP} and {F150LP - F165LP} images is similar,  subtracting the $H_2$ image (i.e., {F140LP - F150LP}) scaled as indicated above from the     {F150LP - F165LP} image should eliminate the $H_2$ emission of the    {F150LP - F165LP} image so that only atomic (mainly C~{\sc iv}) emission  should remain. Given that our STIS long-slit spectrum does not indicate an evolution of the $H_2$ lines with position along the jet and that one fluorescence route often contributes to both images, the error induced by a different spatial morphology of the $H_2$ emission above and below about $1500\,$\AA{} is likely to be negligible.
This scaling, however, produces negative residuals\footnote{This finding is unrelated to the subtraction of the {F165LP} image.} which we attribute to an overestimation of the $H_2$ counts. To avoid strong negative residuals, we use a scaling of 1.2, which reduces the negative residuals while preserving an impression of the problematic region.  The count-rate of the C~{\sc iv} image
is $1.7\pm0.25$ times higher than expected based on the C~{\sc iv} luminosity in our STIS G140M spectrum (the error corresponds to a 10\,\% change of the $H_2$ scaling). We note that this image probably includes a small ($\lesssim 30\,$\%) contribution from other atomic lines. The remaining discrepancy might be due to time variability or cross-calibration problems.

\subsection{Alignment of the images}
The location of the image on the detector depends on which filter is chosen, because the detector and the filters are tilted with respect to the chief ray. Therefore, we took special care aligning the individual filter images as described in the following.

We applied the known filter shifts given in the instrument handbook \citep[Sect. 5.5.3, ][]{STIS} to the F140LP, F150LP, and F165LP images. These shifts are accurate to within about one pixel (30\,mas).
There are no point sources that can be used as references in the field of view and DG Tau itself is invisible shortwards of $\sim1600\,\AA$ where most of the counts in the {F140LP} and    {F150LP} images originate. Therefore, the alignment of the images cannot be improved without additional assumptions. 

For the alignment perpendicular to the jet axis we assumed that the emission close to DG~Tau is symmetric around the jet axis so that the emission centroid perpendicular to the jet axis should match in all filters.  By fitting Voigt profiles to the region  within 125\,mas around the brightest part of the emission we aligned the centroids with shifts up to about 0.5\,pixel (15\,mas). 

The alignment along the jet axis was done assuming that photons with relatively long wavelengths dominate the emission close to DG~Tau and match the images on the side of DG~Tau, while allowing them to differ in the forward jet direction (effects caused by the Point Spread Function (PSF) are only a few percentage points and thus negligible). 
The required shifts are 0.25 and 0.5 pixel (7.5 and 15\,mas), respectively.

We checked these shifts by comparing the spatial count distribution in the    {F150LP - F165LP} image with our STIS G140M spectrum obtained five months before the ACS images. 
This STIS G140M observation covers only the wavelength range from 1514\,\AA{} to 1567\,\AA{}, while most of the detected photons originate from a much larger wavelength range (approximately from $1500$ to $1650\,$\AA). However, the emission is dominated by $H_2$ emission in the F140LP and F150LP images (see Sect.~\ref{sect:origin}). Therefore, we scaled the $H_2$ emission seen with the G140M as suggested by the IUE spectrum (see Sect.~\ref{sect:origin}) and compared the  projections from the long-slit spectrum ($H_2$ and C~{\sc iv} emission) and from the images along the jet axis. These projections agree very well and we estimate an error of $0\farcs05$ for the spatial zero-point. Similarly shifting the {F165LP} and {F150LP} images against each other results in a reduced agreement between the projection and the long-slit data. A proper-motion velocity of 300\,km\,s$^{-1}$, which is typical for optical knots in the DG~Tau jet, would result in an observed shift of  0\farcs13 between the STIS spectrum and the ACS images. However, the dominating $H_2$ emission is about three to four times slower \citep{Schneider_FUV} so that a maximum offset of about $0\farcs03$-$0\farcs04$ is expected, i.e., comparable with the registration accuracy. 
After this procedure the peak in the    {F165LP} is 6\,mas (0.2\,pixel) downstream of the estimated DG~Tau position.

\subsection{Construction of the $H_2$ and C~{\sc iv} images}
Direct subtraction of the {F150LP} from  the    {F140LP} image results in unphysical negative residuals close to DG~Tau which cannot be avoided by changes of the image registration. Inspection of the spatial extent of the emission perpendicular to the jet axis shows that the emission in the {F140LP} filter is more extended than in the    {F150LP} filter; the FHWM perpendicular to the jet axis is approximately $29$\,\% larger. According to synthetic Tiny Tim\footnote{http://www.stsci.edu/hst/observatory/focus/TinyTim} models, differences between the two filters are only 2\,\%.
While the difference in the spatial extent might be partly physical, the negative residuals clearly demonstrate that there are additional instrumental artefacts. We checked that neither thermal settling (the second    {F140LP} image shows a larger FWHM than the first    {F140LP} image) nor spacecraft pointing anomalies (the \texttt{jitter}-files show no anomalies) can explain this property. 
Inspection of calibration images shows that the spatial sizes of point-sources agree reasonably with the Tiny Tim expectation; however, some show differences between the two filters almost as large as those found for the spatial extent perpendicular to the jet axis in our images. Therefore, we had to smooth the {F150LP} and    {F165LP}\footnote{The    {F165LP} needs to be smoothed as well to allow a comparison with results obtained from the    {F140LP} image.} images with a Gaussian of 32.5\,mas to facilitate the construction of the difference images. We checked that this procedure does not significantly change the spatial shape of the emission in the difference images.

\subsection{Deconvolution of the $H_2$ image}
The  $H_2$ image shows spatial structure on the same scale as the PSF. To improve this image, we created an artificial PSF for the $H_2$ image using the mean Tiny Tim PSF of the    {F140LP} and    {F150LP} images smoothed by the same Gaussian as the    {F150LP} data for the difference image. We do not deconvolve the individual images, because those deconvolved images already include different artefacts at similar iteration levels which would be transfered into the difference images. We think that deconvolving the difference image provides a better solution in our case. We note that the Tiny Tim PSFs for the    {F140LP} and    {F150LP} images differ only slightly for the spectrum at hand. We used the Richardson-Lucy deconvolution with 50 iterations implemented by \texttt{arestore} in the CIAO tools\footnote{see \texttt{http://cxc.harvard.edu/ciao/intro/tools.html}} \citep{CIAO}.

\section{Molecular hydrogen: results\label{sect:H2results}}

\begin{figure}
\centering
\fbox{\includegraphics[width=0.48\textwidth]{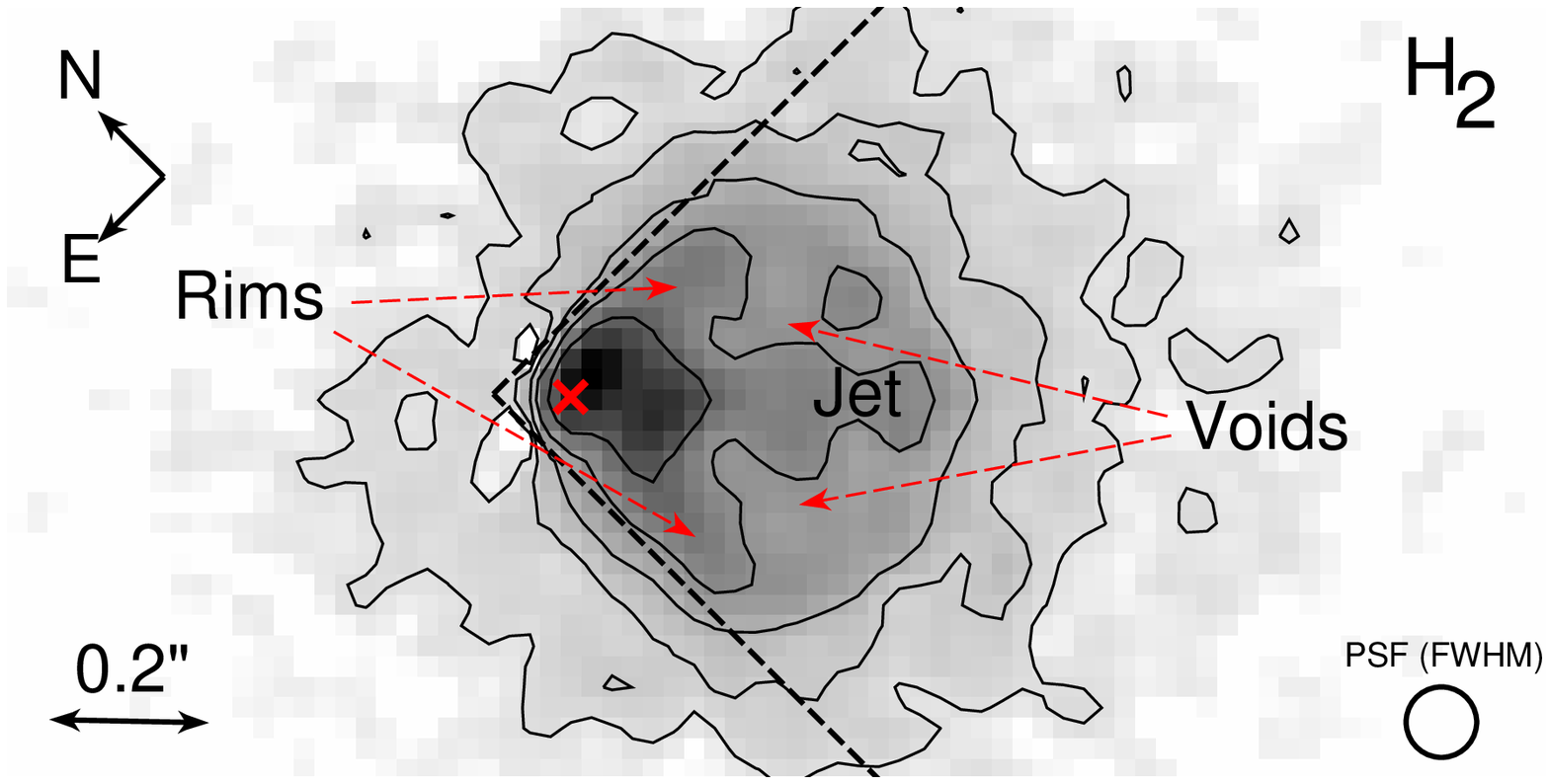}}
\fbox{\includegraphics[width=0.48\textwidth]{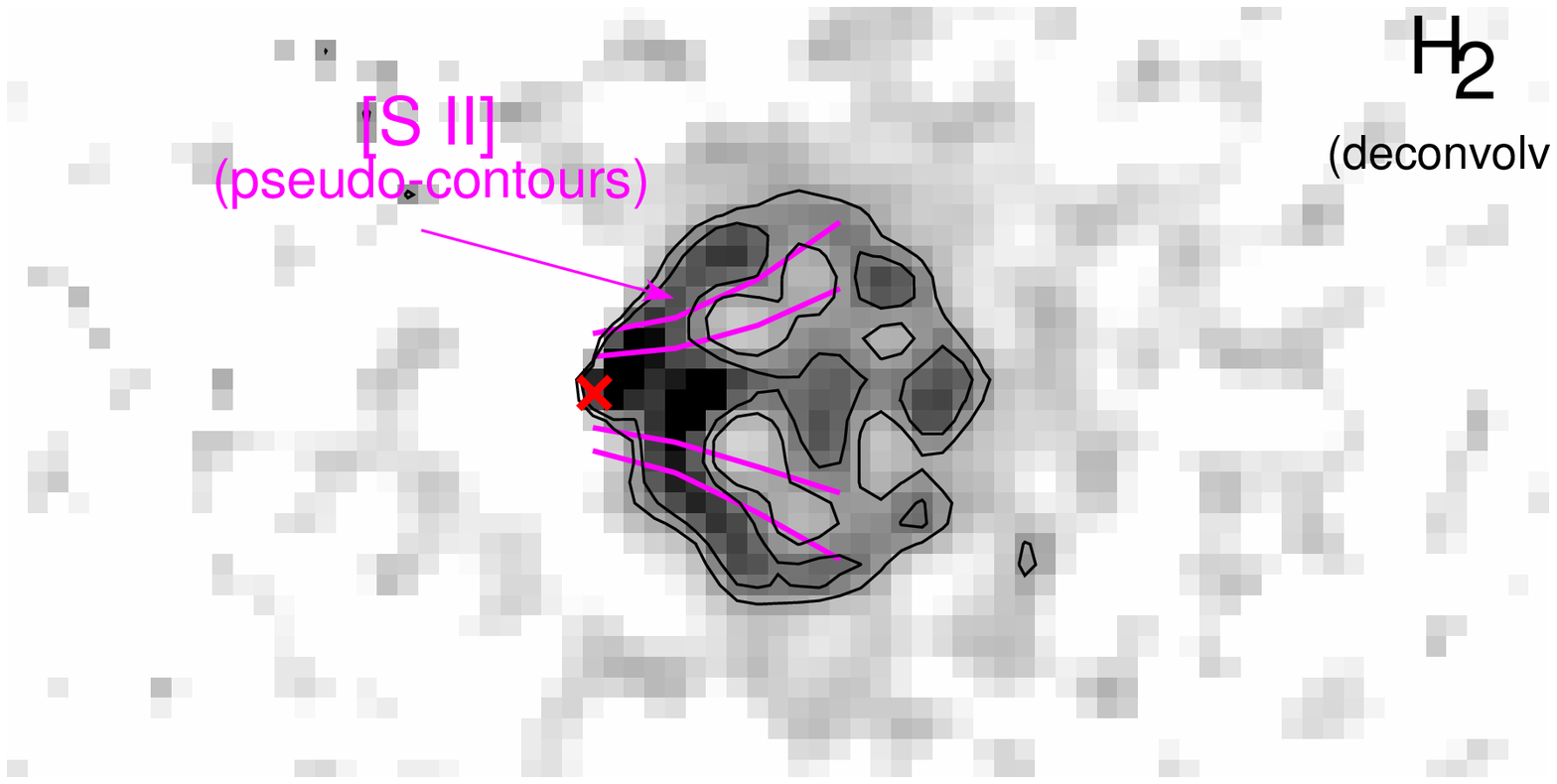}}
\caption{{\bf Top: } Inner part of the $H_2$ emission. Contours start at 10\,counts per pixel increasing by a factor of two. The dashed lines indicate the cone used to extract the radial dependence of the count rate.  
{\bf Bottom: } Deconvolved $H_2$ image. The magenta lines indicate the region containing 68\,\% and 90\,\% (outer lines) of the  [S~{\sc ii}]  emission in the velocity range +25 to -120 km\,s$^{-1}$ perpendicular to the jet axis. }
 \label{fig:H2}
\end{figure}

The $H_2$ images obtained by the procedures described above are shown in Fig.~\ref{fig:H2}.
\subsection{Spatial structure}
The $H_2$ emission consists of a pronounced bright rim, emission along the central jet axis distinct from the rim, and bright emission close to DG~Tau. 
The spatial morphology resembles an outflow-cavity wall structure; however, we argue in Sect.~\ref{sect:relation} that the void in $H_2$ emission is probably filled with other outflowing material so it is not a cavity. Consequently, we speak of the rims when dealing with the V-shaped feature. On the other hand, it is possible  that the DG~Tau jet will develop into a jet-cavity wall structure at larger distances from DG~Tau. Since we do not have data on this region, we will focus on the inner part of the jet/outflow.

We find an opening angle of about 90$^\circ$ between the two rims. Following the rim towards the DG~Tau position, both edges intersect very close to the expected DG~Tau position. From visual inspection, we find that the rims would intersect a flat disk within a few AU from DG Tau.
The rim has a length of approximately $0\farcs3$ (projected 42\,AU).
Comparing the projection perpendicular to the rim in the F140LP, which is dominated by $H_2$ emission with a Tiny Tim model of a linear feature, we
find that the spatial distribution of the emission close to DG Tau resembles the PSF more closely than it does farther away.
This indicates that the (projected) sharpness of the rim decreases with increasing distance to DG~Tau.  
Overall, the emission decreases beyond a radial distance of approximately $0\farcs45$. The outer (downstream) edge and the large structure appear almost circular. A circle of radius $0\farcs25$ encloses more than two thirds of the emission.

\subsection{Outflow direction}
To determine the jet center as a function of distance to the source, we fit Gaussians to slices extending 0\farcs15 along the jet axis. Since the difference images might contain artificial effects due to registration errors, we use the F140LP, F150LP, and F165LP images directly. 
The fits were performed using the emission within 0\farcs5 from the iteratively determined jet axis and the results are shown in Fig.~\ref{fig:rawIm}. 
Within the inner 2\arcsec{} the signal is sufficient for this procedure and a straight jet gives a PA of $224.6\pm0.2^\circ$ ($\chi_{red}\approx1.0$) compatible with previous estimates \citep[e.g., 226$^\circ$; ][]{Solf_1993}.

There is also some asymmetry within the jet; for example, extracting the counts between 0\farcs1 and 0\farcs3 at both sides of the jet center, we find the northwestern side slightly brighter than the southeastern at the 10\,\% level. Although these asymmetries are small, they show that inhomogeneities within the outflow are already generated  close to the source before interactions with the ambient medium take place. Furthermore, the locations of the jet centers show an additional substructure which looks like wiggling (cf. red line in Fig.~\ref{fig:rawIm}). Although the statistical error on each data point is approximately similar to its distance to the straight axis, the offsets are consistent between the filters and so are probably not purely statistical. The half opening angle of the potential wiggling amplitude is below two degrees. Given the substructure of the outflow, the small variations of the jet center probably result from the slight asymmetries in the emission rather than from a change in the outflow direction.

\subsection{Comparison with near-IR $H_2$ data of DG~Tau}
Indications for the V-shaped $H_2$ feature are also seen in near-IR integral field data of DG~Tau obtained with Keck \citep[observation date 2005 Oct 26,][]{Beck_2008}. These authors observed a variety of near-IR $H_2$ lines and in particular present an image of the outflow in the \hbox{$v=1-0\,S(1)$} line at 2.12\,$\mu$m. A similar spatial distribution was probably also present in the Subaru data of \citet{Takami_2004}, since convolving the emission within the inner 0\farcs5 of the forward jet with a Gaussian approximating their seeing conditions results in a Gaussian FWHM of 0\farcs5, which is quite similar to the reported 0\farcs6. Thus, the $H_2$ emission appears stationary over almost one decade.  A similar V-shaped structure of near-IR $H_2$ emission has also been reported for HL~Tau by \citet{Takami_2007}.

\subsection{Relation to the optical jet \label{sect:relation}}
We compare the $H_2$ with low-temperature plasma ($T\sim10^4\,$K) as traced by [S~{\sc ii}] $\lambda6731$ emission  in Fig.~\ref{fig:H2} (bottom). The low-velocity part of the [S~{\sc ii}] emission appears to be located  close to DG~Tau at all times, i.e., within the innermost 0\farcs2 or so \citep{Lavalley_1997, Schneider_FUV}.  To check the spatial relation of both components we compare the distribution of the $H_2$ emission with an artificial image of the low-velocity [S~{\sc ii}] emission (integrated between +25 and -120\,km\,s$^{-1}$) obtained by combining seven consecutive HST long-slit spectra. This [S~{\sc ii}] image covers the innermost $\pm\approx0\farcs25$ perpendicular to the jet axis. Its spatial resolution is about 0\farcs1. Details of the data processing are provided by \citet[][]{Bacciotti_2000, Bacciotti_2002} and \citet{Maurri}. Since this optical image has a lower spatial resolution  than the FUV data, we fit Gaussians to the [S~{\sc ii}] emission perpendicular to the jet axis to estimate the spatial extent. Figure~\ref{fig:H2} shows the 1-$\sigma$ width of the Gaussians corrected for our estimate of the instrumental PSF width\footnote{The shape of the PSF of this artificial image is not well known. Here, we use the width of the Gaussian fitted to a STIS image of a point source ($\sigma=37$\,mas). However, the spatial extent of the continuum along the jet direction in the G750M spectra is $\sigma=50\,$mas,   and so the [S~{\sc ii}] emission might be slightly more concentrated than shown in Fig.~\ref{fig:H2}, bottom.}, i.e., the region containing 68\,\% of the [S~{\sc ii}] emission. We find that the $H_2$ emission extends to larger jet radii than the [S~{\sc ii}] emission (see Fig.~\ref{fig:H2}, bottom), i.e., the [S~{\sc ii}] emission is located between the $H_2$ rims. 
The initial opening angle of the [S~{\sc ii}] emission is also smaller than that of the $H_2$ emission. 
However, beyond about 0\farcs4 from DG~Tau the [S~{\sc ii}] emission extends to larger jet radii than the FUV $H_2$ emission. This might be related to the bubble-like feature seen in the [S~{\sc ii}] data.

\section{Molecular hydrogen: Discussion and interpretation \label{sect:H2discussion}}
Most of the observed FUV $H_2$ emission lines are fluorescently excited by Ly$\alpha$ photons\footnote{A small fraction of the $H_2$ lines contributing to our FUV images might be pumped by other lines, e.g., C~{\sc iv}.} which requires $H_2$ temperatures around 2000\,K. Therefore, the FUV $H_2$ emission traces the amount of warm $H_2$ {\it and} the local Ly$\alpha$ radiation field. In CTTS, the $H_2$ emitting  gas can be located in the outflow or in the warm, upper layers of a flared disk \citep{France_2012}. In the latter case, the dominant pumping source for the fluorescent $H_2$ emission is the strong stellar Ly$\alpha$ emission. 

\subsection{Outflow origin of the FUV $H_2$ emission}

To interpret the observed spatial distribution of the $H_2$ emission (see Fig.~\ref{fig:H2}), we must investigate whether the two voids are due to absorption, or whether they are due to a lack of $H_2$ emission  at intermediate distances to the jet axis. The $H_2$  emission feature close to the jet axis does not provide additional insight, because $H_2$ might be reformed behind an internal shock front within the outflow \citep{Raga_2005}.

For a standard gas-to-dust ratio, a column density equivalent of $N_H=7.6\times10^{20}\,$cm$^{-2}$ is required for $\tau_{FUV}=1$  using the \citet{Fitzpatrick_1999} FUV extinction law averaged between 1400\,\AA{} and 1650\,\AA{} 
and $N_H = 1.8\times10^{21} A_V$\,cm$^{-2}$ \citep{Predehl_1995}. This is compatible with the attenuation of the hot, X-ray emitting plasma which is probably located close to the jet axis at distances of 30-40\,AU from DG~Tau \citep[$N_H\approx1.1\times10^{21}\,$cm$^{-2}$;][]{Guedel_2007}. 
On the other hand, this column density requires an average jet density of $n_H\gtrsim10^6\,$cm$^{-3}$ assuming a circular jet cross-section with a diameter of 0\farcs3 (42\,AU$=6.3\times10^{14}\,$cm) at a distance of 0\farcs2 (40\,AU) from DG~Tau. At this position, we estimate a density of $n_H\lesssim2\times10^6\,$cm$^{-3}$  for the high-velocity jet component by extrapolating the measurement of \citet{Coffey_2008} assuming a conical jet. The lower velocity jet components are located at larger distances to the jet axis and their density is lower by a factor of two. Therefore, the  density required for $\tau_{FUV} = 1$ already appears to be high since a volume filling factor of about unity would be required for the estimated densities which appears unlikely for stellar jets. We note that we neglect $H_2$ line absorption, because the material along the line-of-sight differs in  velocity due to the onion-like velocity structure of the jet with faster streamlines closer to the jet axis \citep{Bacciotti_2000}.

This interpretation is compatible with the observed near-IR lines assuming that they originate in close spatial proximity of the FUV $H_2$ lines (which their similar shape and their approximately similar velocities suggest, see below).  The absorption in the near-IR is strongly reduced compared to the FUV. For example, the transmission is three times larger for the 2.12\,$\mu$m $H_2$ line than for the FUV lines assuming $E(B-V)=0.18$, i.e., the estimated extinction of the hot, X-ray emitting plasma close to the jet axis. This ratio increases for larger column densities. Therefore, the 
V-shape of the near-IR $H_2$ emission also indicates a lack of intrinsic $H_2$ emission at small and intermediate distances from the jet axis.

With this interpretation, the voids in the $H_2$ emission originate from a lack of warm $H_2$ ($T\approx 2000\,$K) or from a lack of pumping Ly$\alpha$ photons at the location of the warm $H_2$ gas.
The models for molecular outflows by \citet{Panoglou_2012} suggest that $H_2$ survives mainly in the outer layers of a disk wind and that the temperatures of the streamlines launched at 1\,AU are in the required range ($\sim2000\,$K). Streamlines launched closer to the star have a lower $H_2$ abundance and higher temperatures which can explain the lack of (FUV) $H_2$ emission at smaller distances to the jet axis. In this scenario, the decreasing sharpness of the rim edges is a natural consequence of an optically thin cone with an increasing wall-thickness.

The wide opening-angle close to the source resembles the bubble-like features already discussed by \citet{Bacciotti_2000}.
Comparing the FUV $H_2$ emission with the optical jet traced by [S~{\sc ii}]\,$\lambda6731$ emission ($T\sim10^4\,$K, cf. Fig.~\ref{fig:H2}, bottom), we find that the distance of the rim to jet axis is about twice as large as the 1-$\sigma$ width of  the [S~{\sc ii}] emission. 
The mean velocity of the [S~{\sc ii}] emitting material shown in Fig.~\ref{fig:H2} is $-40$ to $-70$\,km\,s$^{-1}$ (higher velocities pertain to the regions closer to the jet axis) which is about twice the velocity of the FUV $H_2$ lines  \citep[$-10$ to $-30$\,km\,s$^{-1}$, GHRS; $-9.2\pm0.9$\,km\,s$^{-1}$, STIS E140M; $-27$\,km\,s$^{-1}$, G140M; $-20$ \dots $-30$\,km\,s$^{-1}$; see ][respectively]{Ardila_2002, Herczeg_2006, Schneider_FUV}. The FUV $H_2$ velocities are comparable with those of the near-IR $H_2$ lines \citep[$-20$\,km\,s$^{-1}$, $-2\pm18$\,km\,s$^{-1}$, and $ -9$\,km\,s$^{-1}$; see][respectively]{Takami_2004,Beck_2008,Greene_2010}.  According to \citet{Ferreira_2006}, the poloidal velocities are inversely proportional to the launching radius for similar magnetic lever arms, which fits perfectly considering that the extent of the $H_2$ emission perpendicular to the jet axis is about twice as large as that of  the [S~{\sc ii}] emission. 

Another tracer of the molecular outflow is CO line emission \citep[e.g.,][]{Herczeg_2011} that has also been detected in the DG~Tau outflow by \citet{Podio_2012}. 
In CO,  V-shaped morphologies are quite common on larger scales \citep[e.g.,][]{Arce_2006} and can be interpreted as the interaction of a wide opening-angle outflow with the envelope \citep{Offner_2011}. The expected opening-angle of such a wind is comparable to the measured opening-angle of $90^\circ$ for DG~Tau and we speculate that the shocks develop because of the interaction of the wide opening-angle wind with the extended atmosphere of the disk or with the innermost remnants of the envelope. This might also explain the stationary appearance of the rims.

The CO emission is unresolved in Herschel/PACS observations \citep{Podio_2012}, but the large CO luminosity and high-excitation lines argue for excitation in C-type shocks and against an origin in the disk. In general, typical shock velocities required to excite the CO emission are \hbox{a few 10\,km\,s$^{-1}$} while measured line-shifts are often only a few km\,s$^{-1}$. Recently \citet{Podio_2013} found a blue-shifted wing of CO emission with velocities of a few km\,s$^{-1}$ for DG~Tau. Those CO velocities are factors of a few below the $H_2$ velocities, so that extrapolating our $H_2$ vs [S~{\sc ii}] $\lambda6731$ results, we expect the CO to be located outside the $H_2$ cone.

We consider an origin of the FUV $H_2$ emission in the disk atmosphere to be unlikely. First, it cannot explain the observed blue-shifted velocities, because no line shifts are expected due to (Kepler-) rotation. Second, it requires an unlikely massive outflow, because the emission from the far-side of the disk needs to be absorbed by the outflow filling the region above the disk surface. Third, it requires that the faster FUV $H_2$ emission be located at larger distances from the jet axis than the slightly slower near-IR $H_2$, because the near-IR lines are excited by shocks which are unlikely to occur deeper within the disk than the fluorescent FUV $H_2$ emission. Therefore, all observed properties of the FUV $H_2$ indicate a disk wind origin.

\subsection{The pumping source}
\begin{figure}
\centering
\includegraphics[width=0.48\textwidth]{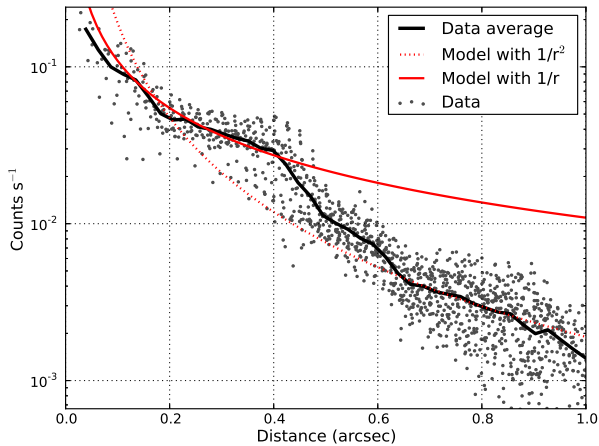}
\caption{Count-rate as a function of distance from the nominal DG~Tau position. The data have been extracted within the 90$^\circ$ cone indicated in Fig.~\ref{fig:H2}. The two models differ in their normalization. \label{fig:radial} }
\end{figure}

The FUV $H_2$ emission requires continuous (Ly$\alpha$) pumping, because the lifetime of the upper levels is about 14 orders of magnitude shorter than that of the near-IR ro-vibrational transitions ($10^{-8}$ vs  $10^6$\,s). There are different possible sources for pumping Ly$\alpha$ photons: DG~Tau itself, Ly$\alpha$ emission from fast shocks within the atomic jet illuminating the cone of warm $H_2$, and low-velocity shocks almost co-spatial with the $H_2$ emission.  
The ratios of the FUV $H_2$ lines do not change  along the jet axis arguing for a similar excitation emission close to DG~Tau and farther out. However, the innermost region around DG~Tau is invisible so that the on-source spectrum might be different as in the case of T~Tau \citep{Walter_2003, Saucedo_2003}. On the other hand, we would expect that, in a first order approximation, the surface luminosity drops approximately as $\sim r^{-2}$ with  distance $r$ from DG~Tau, if the stellar FUV flux is the dominant excitation source. 
Figure~\ref{fig:radial} shows the radial dependence of the count-rate. At larger  distances, the $r^{-2}$ model reasonably describes the data, but for $r\lesssim0\farcs4$ the model with $r^{-1}$  provides a better description of the data. The largest deviation from the $r^{-2}$ model is caused by the knee corresponding to the rims and the jet-like feature  ($r=0\farcs2\dots0\farcs4$, a projection along the rims also shows a similar excess in the knee region). 
This is also present in the data analyzed by \citet{Beck_2008} at the same location. It is therefore unlikely that the spatial $H_2$ luminosity distribution is caused by inhomogeneties of the Ly$\alpha$ radiation field  since the near-IR $H_2$ emission is probably shock excited \citep{Beck_2008, Takami_2004} and not pumped by FUV photons.
However, we note that the outflow momentum argument of \citet[][]{Takami_2004} might overestimate the momentum, because the V-shape of the emission was not known at that time. The V-shape provides a larger illuminated surface area for the UV photons and a single streamline can be excited multiple times (similar to the emission from multiple, unresolved shocks within an outflow). Therefore, less outflowing mass would be required compared to a single dense clump. Thus, we consider the measured line ratios of \citet{Beck_2008} to be the strongest indication for shock excitation of the near-IR lines. In any case, we favor a scenario in which the spatial distribution of the $H_2$ emission is dominated by the amount of warm $H_2$ and not by inhomogeneities in the Ly$\alpha$ radiation field. 

We now investigate whether  the observed FUV H$_2$ emission can be pumped by  the FUV flux associated with the high-velocity shocks ($v_{shock}\gtrsim100$\,km\,s$^{-1}$) close to the jet axis traced by C~{\sc iv} $\lambda\lambda1548,1551$, and X-ray emission, or by low-velocity shocks ($v_{shock}\sim$ 30\,km\,s$^{-1}$) that are probably farther away from the jet axis.
\citet{Curiel_1995} found that the conversion of Ly$\alpha$ flux to $H_2$ fluorescence emission might be relatively efficient (about $15\,$\%). The total $H_2$ flux (see Sect. \ref{sect:origin}) would therefore require a Ly$\alpha$ luminosity in excess of $2\times10^{-12}$\,erg\,s$^{-1}$\,cm$^{-2}$. Shock models for high velocity shocks \citep[$v_{shock}\gtrsim150$\,km\,s$^{-1}$, ][]{Hartigan_1987} predict Ly$\alpha$ luminosities $\lesssim10$\,times higher than C~{\sc iv} luminosities. Thus, the  Ly$\alpha$ luminosity associated with these high-velocity shocks is, based on the observed C~{\sc iv} luminosity,   insufficient to power the $H_2$ emission by about an order of magnitude as $L_{\text{C\,{\sc iv}}} \approx 0.1 L_{H_2}$. For lower velocity shocks, the ratio between Ly$\alpha$  and C~{\sc iv} emission increases, so that slower shocks can provide a sufficient Ly$\alpha$ photon flux without increasing the emission from higher temperature plasma tracers such as C~{\sc iv}. These lower velocity shocks would be visible in [O~{\sc i}]\,$\lambda6300$. Using the \citet{Hartigan_1987} models, we estimate that the Ly$\alpha$ luminosity is more than about 30 times the [O~{\sc i}]\,$\lambda6300$ luminosity. Assuming that the [O~{\sc i}] emission is dominated by shocks with a velocity of about 30--100\,km\,s$^{-1}$, only a small percentage of the Ly$\alpha$ photons are needed to pump the fluorescent $H_2$ emission for the detected  luminosity. We therefore favor these low velocity shocks as the origin of the pumping Ly$\alpha$ photons.

\section{The intermediate temperature jet traced by C~{\sc iv} emission: Results and discussion \label{sect:c2}}
\begin{figure}
\vspace*{0.2cm}
\centering
\fbox{\includegraphics[width=0.48\textwidth]{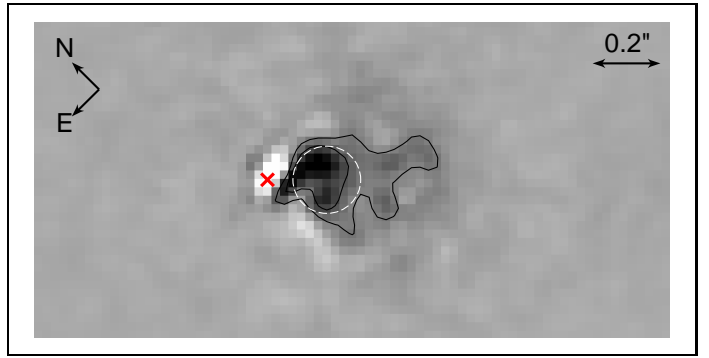}}
\caption{The inner part of the C~{\sc iv} emission. The circle shows the approximate region of the C~{\sc iv} emission seen in the STIS long-slit spectrum. \label{fig:Civ} }
\end{figure}

Our C~{\sc iv} image is shown in Fig.~\ref{fig:Civ}. Although this image is affected by some reduction artefacts, it demonstrates that the majority of the emission falls within a circle about 0\farcs2 from DG~Tau with $r=0\farcs1$, i.e., compatible with the region where the C~{\sc iv} is seen in the STIS long-slit spectrum \citep{Schneider_FUV}. 

\citet{Schneider_FUV} discuss some scenarios resulting in stationary C~{\sc iv} emission including magnetic heating, a hot inner wind (e.g., a stellar wind) and strong shocks within the jet, possibly caused by collimation.
Another interesting scenario is that instabilities  within the jet lead to a number of small, unresolved shocks. In the models of \citet{Staff_2010},  the plasma temperature close to the jet axis  is a few times $10^5\,$K. The heating of this high-temperature plasma is caused by numerous smaller shocks due to instabilities, e.g., kinks and wobbles within the jet.
The jet is initially stabilized by magnetic fields, but beyond a certain point instabilities develop. The corresponding region is probably close to the Alfv\'en surface explaining that collimation is observed at similar scales. This could explain the observed offset between the peak of the hot plasma and the stellar position and that the high-temperature plasma is constantly located at the same distance from the source.

\section{The outer knots: Results and discussion \label{sect:c3}}
The outer knots are visible in all three filter images (see Fig.~\ref{fig:rawIm}). 
The distances of the knots compare well with the positions observed in the optical STIS spectrum (7\farcs2 and 13\farcs2)\footnote{For a typical knot velocity in the DG~Tau jet of 300\,km\,s$^{-1}$, the proper-motion between the two observations is about 0\farcs1.} indicating that the FUV emission largely overlaps with the optical jet seen in [S~{\sc ii}]\,$\lambda6731$, for example. 
The knot at 7\arcsec{} does not show a bow-like structure but appears rather circular. Nevertheless, regions of higher surface brightness tend to be located slightly closer to the shock front. 
The count-rates of the outer knots integrated over a circle of 2\,arcsec are only about 5\,\% of that of the inner emission component, but the region of the STIS long-slit spectrum contains less than 1\,\% of the counts observed in a similar region close to DG~Tau, which explains why they were not detected in the STIS long-slit spectrum. The integrated count ratios between the filters differ from the ratios obtained close to DG~Tau so that we cannot constrain the spectral origin of the emission.

The outer FUV emission is located farther away from DG~Tau than the resolved outer X-ray knot \citep{Guedel_2008, Guedel_2011}, which has an expected position during the observations of about $6\pm0\farcs5$. 
This offset between the X-ray and FUV emission is incompatible with a scenario  where  the FUV  {\it and} the X-ray emission originate from the same (forward) shock. A reverse facing shock might explain this pattern; however, we regard this to be unlikely since both  the optical and X-ray knots  move with rather similar velocities into the forward jet direction. On the other hand, it is possible that the X-rays trace the jet shock/Mach disk. Depending on the density contrast between the medium in front of the fast jet and that of the fast jet itself, bow shock and Mach disk have different shock velocities. If the medium in front of the jet is denser than the jet itself, the velocity of the Mach disk is higher than that of the bow shock \citep{Hartigan_1989}. 

Alternatively, the X-ray emitting plasma could have been heated a few years before the X-ray observations, since its cooling time is rather long, several years according to \citet[][]{Guedel_2008}. Thus, the X-ray emitting plasma might be the reminiscence of older shocks, which are not visible in lower temperature tracers any longer because of the higher radiative cooling rate at lower temperatures (an order of magnitude for $T\sim10^5$\,K at the same density).

\section{Summary and conclusions \label{sect:sum}}
We present high spatial resolution FUV images of the DG~Tau jet and construct images tracing $H_2$ and C~{\sc iv} emission. The $H_2$ image shows that the emission is concentrated into the forward jet direction. The $H_2$ emission consists of a cone with two bright rims which are slightly asymmetric. They are unresolved close to DG~Tau and reach out to a radial distance of about 50\,AU. Additionally, there is $H_2$ emission close to the jet axis. The bright rims are also seen in near-IR $H_2$ lines with a  very similar spatial morphology \citep{Beck_2008}. We conclude that the origin of the bright rims is enhanced emission close to the surface of a wide opening-angle cone and absorption of the far side of the outflow by material located close to the jet axis.
The stationarity of this feature indicates that the emission is not caused by individual knots but rather is due to ongoing heating by local shocks. These low velocity shocks also provide the Ly$\alpha$ photons required for the FUV $H_2$ emission. Additional fluorescent $H_2$ emission excited by stellar Ly$\alpha$ photons might explain the excess emission seen close to DG~Tau and the overall radial dependence of the $H_2$ emission. Extended disk-wind models by \citet{Panoglou_2012} show that molecular hydrogen can survive in the outer streamlines and the $H_2$ emission is indeed located about twice  as far from the jet axis as the faster [S~{\sc ii}] emission. Therefore, the $H_2$ emission probably traces the outer part of a disk wind and the voids originate from a lack of sufficient $H_2$ interior of the rims.

The C~{\sc iv} image shows that the hot plasma is confined to the region close to the jet axis which was covered by the STIS long-slit spectrum \citep{Schneider_FUV}.
The FUV emission seen at the position of the outer knots 
overlaps with low temperature plasma ($T\sim10^4$\,K) while being offset from the X-ray emission. 

These results are consistent with a picture of the DG~Tau jet, in which the high-velocity part of the jet is launched at small distances from DG~Tau while the low-velocity part of the outflow, i.e., the molecular outflow, is less collimated with an opening angle of about 90$^\circ$ and is probably launched at larger distances from DG~Tau. Its emission and shape is stationary over five years and unrelated to individual knots. Monitoring of the higher temperature plasma is required to check its relation to knots.

\begin{acknowledgements}
We thank Francesca Bacciotti and Lorenzo Maurri for providing the [S~{\sc ii}] image.
PCS was supported by the DLR under grant 50 OR 1112.
The paper is based on observations obtained by the Hubble Space Telescope. Some of the data presented in this paper were obtained from the Multimission Archive at the Space Telescope Science Institute (MAST). STScI is operated by the Association of Universities for Research in Astronomy, Inc., under NASA contract NAS5-26555. Support for MAST for non-HST data is provided by the NASA Office of Space Science via grant NAG5-7584 and by other grants and contracts.
The publication is supported by the Austrian Science Fund (FWF).
\end{acknowledgements}

\bibliographystyle{aa}
\bibliography{dg}

\end{document}